\documentclass[12pt]{iopart}
\usepackage{graphicx}
\begin{document}

\title
[Production of nuclei and antinuclei in pp and Pb-Pb collisions]
{Production of nuclei and antinuclei in pp and Pb-Pb collisions with ALICE at the LHC}

\author{Natasha Sharma, for the ALICE Collaboration}

\address{Department of Physics, Panjab University, Chandigarh, India-160014.}
\ead{natasha.sharma@cern.ch}
\begin{abstract}

We present first results on the production of nuclei and antinuclei such 
as (anti)deuterons, (anti)tritons, (anti)$^{3}\rm{He}$ and 
(anti)$^{4}\rm{He}$ in $\rm{pp}$ collisions 
at $\sqrt{s}$~=~7~TeV and Pb-Pb collisions at $\sqrt{s_{\rm{NN}}}$ = 2.76 TeV. 
These particles are identified using their energy loss (d$E$/d$x$) information in 
the Time Projection Chamber of the ALICE experiment. The Inner Tracking System 
gives a precise determination of the event vertex, by which primary and secondary 
particles are separated. The high statistics of over 360 million events for $\rm{pp}$ and 
16 million events for Pb-Pb collisions give a significant number of light nuclei and antinuclei 
(Pb-Pb collisions: $\sim30,000$ anti-deuterons($\bar{\rm{d}}$) and
 $\sim4$ anti-alpha($^{4}\overline{\rm{He}}$) ). 
The predictions of various particle ratios from the THERMUS model is also discussed.


\end{abstract}



\section{Introduction}

In the recent years, a lot of progress has been made by the heavy-ion collision 
experiments to search for the heavier (anti)nuclei and (anti)hypernuclei.
It is important to study these nuclei and antinuclei in details in terms of 
their yields, spectra, flow etc., to understand their production mechanism in a collision. 
A Large Ion Collider Experiment (ALICE) at LHC  has excellent particle identification 
capabilities using its various subsystems~\cite{Alex}, to study these nuclei and antinuclei
with large statistics data.
The first results are presented for identified nuclei and antinuclei in mid-rapidity 
region for $\rm{pp}$ collisions at $\sqrt{s}$ = 7 TeV and Pb-Pb collisions at $\sqrt{s_{\rm{NN}}}$ =2.76~TeV 
at the LHC. These particles are identified using their specific energy loss (d$E$/d$x$) measurements in 
the Time Projection Chamber (TPC) of the ALICE experiment.

\begin{figure}
\begin{center}
\includegraphics[width=0.60\linewidth]{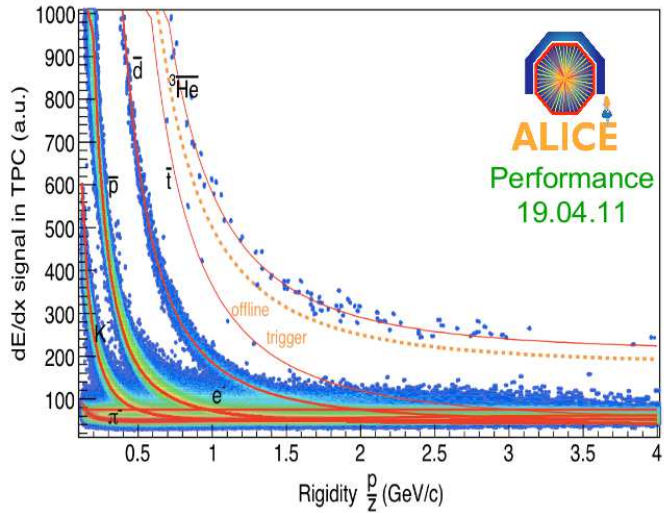}
\caption{(Color online) Specific energy loss d$E$/d$x$ vs. rigidity (momentum/charge) of negatively charged TPC
tracks for Pb-Pb collisions ($\sqrt{s_{\rm{NN}}}$ = 2.76 TeV). The solid lines are 
parametrization of the Bethe-Bloch curve.}
\label{dedx}
\end{center}
\end{figure}

\section{Experiment}
For the present study, over $360$ million triggered events for $\rm{pp}$ 
collisions ($\sqrt{s}$ = 7 TeV) and nearly 16 million triggered events for 
Pb-Pb collisions ($\sqrt{s_{\rm{NN}}}$ = 2.76 TeV) are analyzed.
The various nuclei and antinuclei covered in this analysis are d($\bar{\rm{d}}$), t($\bar{\rm{t}}$), 
$^{3}\rm{He}$($^{3}\overline{\rm{He}}$), and $^{4}\rm{He}$({$^{4}\overline{\rm{He}}$).
We have used the Time Projection Chamber (TPC)  which has full azimuthal
acceptance for tracks in the pseudo-rapidity region 
$|\eta|$ $<$ 0.9. 
The specific energy loss (d$E$/d$x$) versus rigidity (momentum/charge) 
of the negatively charged TPC tracks is shown in Fig.~\ref{dedx}. Antinuclei 
are clearly identified over the wide range of momenta. 
The Inner Tracking System (ITS) containing six silicon layers, is used for precise determination of the event vertex, 
by which primary and secondary particles are separated. Primary tracks are selected with the condition that, 
at least one cluster in the ITS is associated to the track. 

Secondary tracks are further rejected using the distance-of-closest approach (DCA) to the reconstructed primary vertex 
position. 
The $\rm{DCA}_{Z}$ distribution (Z-axis is along beam line) of 
identified antinuclei shows a very small number of tracks with $\rm{DCA}_{Z}$ value greater than 1.0 cm. 
A $\rm{DCA}_{Z}$ cut of 1.0 cm is applied in addition to standard track selection cuts, 
which reduces the fraction of secondary and back-scattered nuclei. 

\begin{figure}
\begin{center}
\includegraphics[scale=0.31]{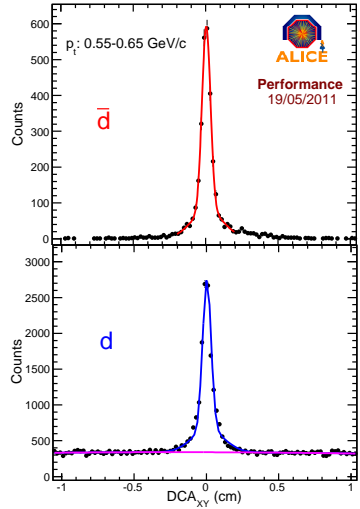}
\caption{$\rm{DCA}_{XY}$ distribution of identified antideuterons (top panel) and deuterons (bottom panel) 
in the transverse momentum range $0.55$ GeV/$c$ $< \ensuremath{p_{\rm t}} < 0.65$ GeV/$c$ 
for Pb-Pb collisions ($\sqrt{s_{\rm{NN}}}$ = 2.76 TeV).}
\label{DCA_plot}
\end{center}
\end{figure}


The probability of antinucleus production by interaction of particles with the detector material is very small. However, nuclei can be produced by interactions with the material.
The final counts of primary d($\bar{\rm{d}}$), t($\bar{\rm{t}}$), and $^{3}\rm{He}$($^{3}\overline{\rm{He}}$) are obtained 
by comparing the $\rm{DCA}_{XY}$ distribution of nuclei and antinuclei in various transverse momentum ($p_{\rm{t}}$) slices. 
Figure~\ref{DCA_plot}
shows an example 
for d and $\bar{\rm{d}}$ in the transverse momentum range $0.55$ GeV/$c$ $<\ensuremath{p_{\rm t}}< 0.65$ GeV/$c$
for Pb-Pb collisions. 
The top panel (for $\bar{\rm{d}}$) 
shows no significant number of tracks in $|\rm{DCA}_{XY}|$ $>$ 1.0 cm region, whereas bottom panel (for d) 
shows 
an almost flat background. The raw yield is obtained from the area under the peak minus background. 
A similar procedure is used to get the raw yields of d($\bar{\rm{d}}$), t($\bar{\rm{t}}$), and $^{3}\rm{He}$($^{3}\overline{\rm{He}}$) 
for $\rm{pp}$ collisions at $\sqrt{s}$ = 7 TeV. 


\section{Results}

The raw spectra of d($\bar{\rm{d}}$), t($\bar{\rm{t}}$), and $^{3}\rm{He}$($^{3}\overline{\rm{He}}$) are obtained
for $\rm{pp}$ collisions at $\sqrt{s}$~=~7~TeV and for Pb-Pb collisions at $\sqrt{s_{\rm{NN}}}$ = 2.76 TeV. 
We observed about 20k antideuterons, 20 antitritons, and 20 $^{3}\overline{\rm{He}}$ candidates for the $\rm{pp}$ collisions 
collected in 2010. For Pb-Pb data, we observed nearly 35k antideuterons, 120 
antitritons, and 700  $^{3}\overline{\rm{He}}$ candidates during the same year. As an example 
the left panel of Fig.~\ref{thermus} shows the raw spectra of $\bar{\rm{d}}$ for $\rm{pp}$ collisions at $\sqrt{s}$ = 7 TeV. 
Antideuterons are identified in the transverse momentum range $0.5$ GeV/$c$ $< \ensuremath{p_{\rm t}} < 1.4$ GeV/$c$.
The right panel of Fig.~\ref{thermus} shows the raw yield of  $^{3}\overline{\rm{He}}$ for Pb-Pb collisions at $\sqrt{s_{\rm{NN}}}$ = 2.76 TeV. 
 $^{3}\overline{\rm{He}}$ are identified in the transverse momentum range $1.2$ GeV/$c$ $< \ensuremath{p_{\rm t}} < 8.0$ GeV/$c$. To get the final yields of nuclei and antinuclei the efficiency correction and annihilation correction have to be taken into account, this work is ongoing.
We have also observed four candidates of $^{4}\overline{\rm{He}}$ in 17.8 million Pb-Pb collisions~\cite{Alex}. This is the 
full statistics of data ALICE has taken during heavy-ion run in the year 2010. 

It will be interesting to compare the final yields of these nuclei and antinuclei in $\rm{pp}$ collisions at $\sqrt{s}$ = 7 TeV and Pb-Pb collisions at $\sqrt{s_{\rm{NN}}}$ = 2.76 TeV and also with 
the RHIC results for Au-Au collisions at $\sqrt{s_{\rm{NN}}}$ = 200 GeV ~\cite{StarNature}.

\begin{figure}
\begin{center}
\includegraphics[width=0.44\linewidth]{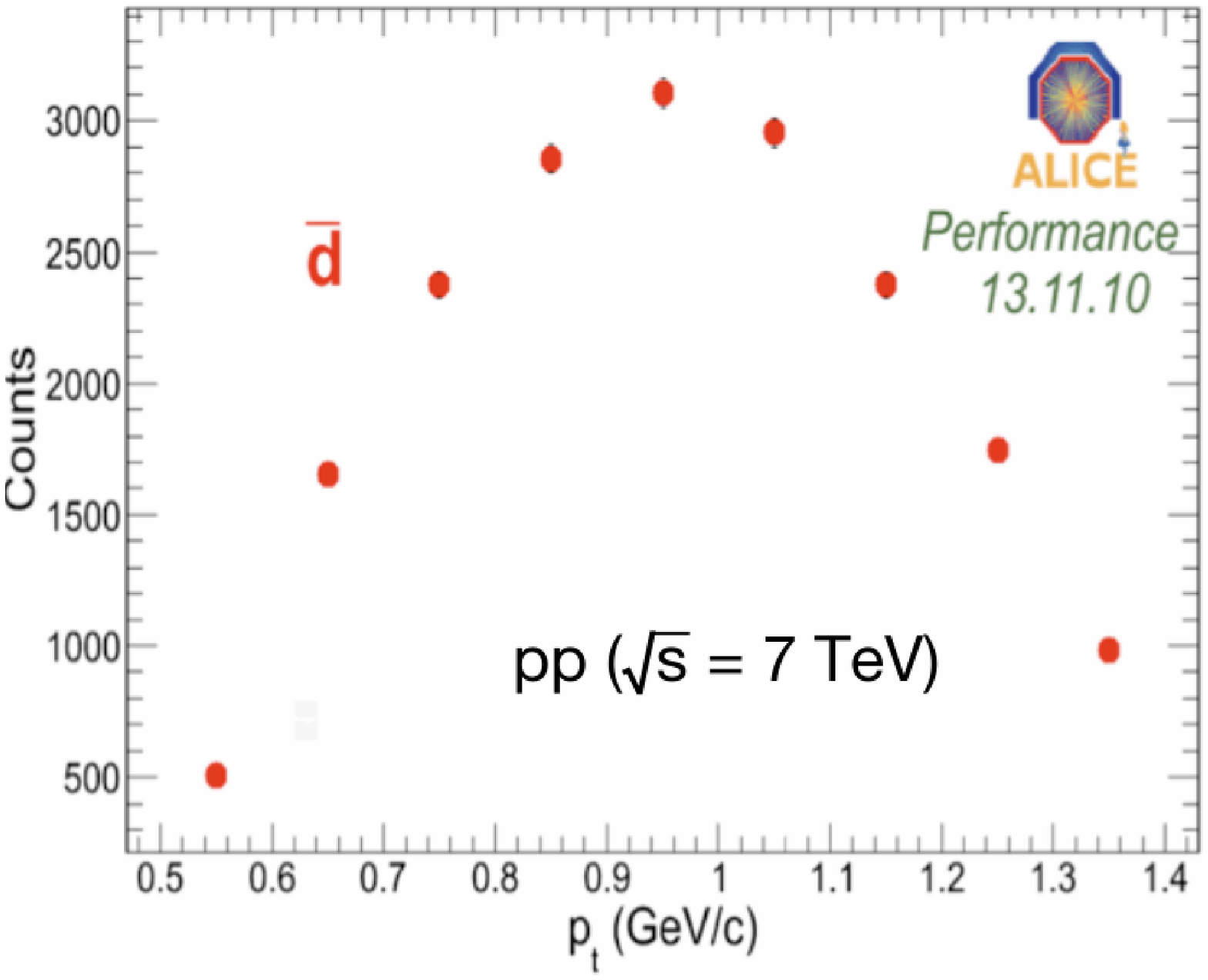}
\includegraphics[scale=0.42]{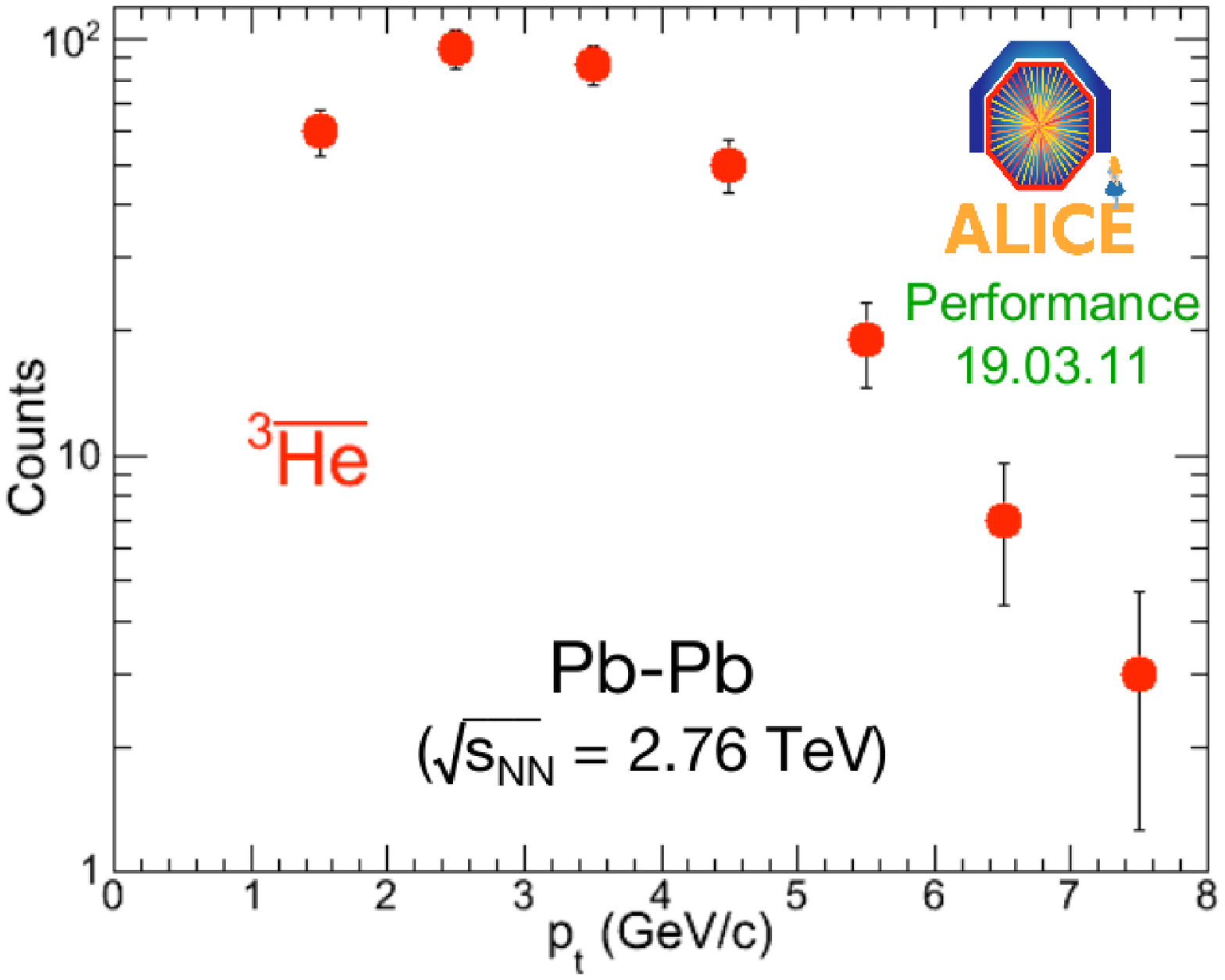}
\caption{Left Panel: Raw yield of antideuterons as a function of transverse momentum ($\ensuremath{p_{\rm t}}$) for $\rm{pp}$
collisions ($\sqrt{s}$ = 7 TeV), only statistical errors are included.
Right Panel: Raw yield of $^{3}\overline{\rm{He}}$ as a function of transverse momentum ($\ensuremath{p_{\rm t}}$)
for Pb-Pb collisions ($\sqrt{s_{\rm{NN}}}$ = 2.76 TeV), 
only statistical errors are included.}
\label{thermus}
\end{center}
\end{figure}

Furthermore, we will compare these ratios with statistical thermal model predictions and coalescence approaches. Figure~\ref{Temp} shows the ratio of particles with different masses, assuming chemical freeze-out temperature ($T$) between 110 MeV and 170 MeV. This shows that the particle ratios are very sensitive to the freeze-out temperature~\cite{ourpaper}. These calculations are performed for Pb-Pb collisions at $\sqrt{s_{\rm{NN}}}$~=~2.76~TeV using the grand canonical approach of the THERMUS code~\cite{Wheaton:2004qb}.

 




\begin{figure}
\begin{center}
\includegraphics[scale=0.33]{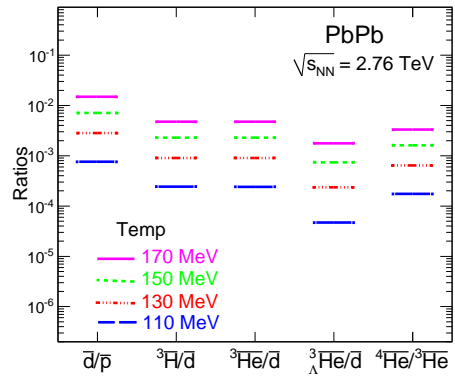}
\caption{Ratios of particles with different masses predicted using the statistical thermal model (THERMUS~\cite{Wheaton:2004qb}) for Pb-Pb collisions at $\sqrt{s_{\rm{NN}}}$ = 2.76 TeV.}
\label{Temp}
\end{center}
\end{figure}

\end{document}